\documentclass[a4paper]{jpconf}
\usepackage{graphicx}
\begin{document}

\title{Close-up of primary and secondary asteroseismic CoRoT targets and the ground-based follow-up observations}

\author{K Uytterhoeven$^1$, E Poretti$^1$, M Rainer$^1$, L Mantegazza$^1$, W Zima$^2$, C Aerts$^{2,3}$, T Morel$^2$, A Miglio$^4$, K Lefever$^2$, P J Amado$^5$, S\,Mart\'{\i}n-Ruiz$^5$, P Mathias$^6$, J C Valtier$^6$, M Papar\'o$^7$,  J M  Benk\H{o}$^7$ and the CoRoT/SWG Ground-based Observations Working Group}

\address{$^1$ INAF-Brera Astronomical Observatory, Via E. Bianchi 46, 23807 Merate, Italy}
\address{$^2$ Institute of Astronomy, KULeuven, Celestijnenlaan 200D, 3001 Leuven, Belgium}
\address{$^3$ Department of Astrophysics, University of Nijmegen, IMAPP, PO Box 9010, 6500 GL Nijmegen, the Netherlands}
\address{$^4$ Institut d'Astrophysique et de G\'eophysique de l'Universit\'e de Li\`ege, Alle\'e du 6 Ao\^ut 17, 4000 Li\`ege, Belgium}
\address{$^5$ Instituto de Astrof\'{\i}sica de Andaluc\'{\i}a (CSIC), Apartado 3004, 18080 Granada, Spain}
\address{$^6$ Observatoire de la C\^ote d'Azur, GEMINI, CNRS, Universit\'e Nice
Sophia-Antipolis, BP 4229, 06304 Nice Cedex 4, France}
\address{$^7$ Konkoly Observatory, P.O. Box 67, 1525 Budapest, Hungary}

\ead{katrien.uytterhoeven@brera.inaf.it}

\begin{abstract}
To optimise the science results of the asteroseismic part of
the CoRoT satellite mission a complementary simultaneous ground-based
observational campaign is organised for selected CoRoT targets. The
observations include both high-resolution spectroscopic and
multi-colour photometric data. We present the preliminary results of
the analysis of the ground-based observations of three targets.  A
line-profile analysis of 216 high-resolution FEROS spectra of the
$\delta$ Sct star HD~50844 reveals more than ten pulsation frequencies in the
frequency range 5--18 d$^{-1}$, including possibly one radial fundamental
mode (6.92 d$^{-1}$). Based on more than 600 multi-colour photometric
datapoints of the $\beta$ Cep star HD~180642, spanning about three years and
obtained with different telescopes and different instruments, we
confirm the presence of a dominant radial mode
$\nu_1=5.48695\,$d$^{-1}$, and detect also its first two
harmonics. We find evidence for a second mode
$\nu_2=0.3017$\,d$^{-1}$, possibly a g-mode, and indications for two
more frequencies in the 7--8 d$^{-1}$ domain. From Str\"omgren photometry we find evidence for the hybrid  $\delta$ Sct/$\gamma$ Dor character of the F0 star HD~44195, as frequencies near 3 d$^{-1}$ and 21 d$^{-1}$ are detected simultaneously in the different filters.  
\end{abstract}

\section{The simultaneous ground-based follow-up observations of CoRoT targets}
The asteroseismic window of the CoRoT satellite mission aims at the
monitoring of several types of pulsators along the Main Sequence. To
optimise the science results, its targets are carefully chosen and
selected. Preparatory observations from the ground have been a key stone
in the selection process \cite{ref1}. With the CoRoT satellite
successfully launched (December 2006), simultaneous ground-based
observations are very important and are complementary to the space
data.  Multi-colour photometry provides colour information, which
allows identification of the degree $\ell$ by means of amplitude ratios and
phase shifts, while high-resolution spectroscopy allows the detection
of high-degree modes and the identification of both the degree $\ell$ and the azimuthal order $m$. 

In this framework a simultaneous ground-based observing campaign was
organised by the CoRoT/SWG Ground-based Observations Working
Group. Large Programme proposals (i.e. guaranteed observing time
during four observing seasons) were submitted and accepted at ESO La
Silla (FEROS/2.2m; 10+5 nights per semester), Observatoire de Haute
Provence (SOPHIE/1.92m; 10+5 nights per semester) and Calar Alto
Astronomical Observatory (FOCES/2.2m; 10+10 nights per semester),
with the aim to obtain multi-site time-series of high-resolution
spectra of a selection of $\delta$ Sct, $\gamma$ Dor, $\beta$ Cep and
Be CoRoT primary and secondary targets. 

The first two observing
seasons (winter 2006-2007 and summer 2007), (nearly) coinciding with CoRoT's IR01 (first
Initial Run) and LRc1 (first Long Run in the center direction), have
been succesfully completed. \Tref{logbook} gives an overview of
the targets and the amount of spectra obtained. 

In addition to the
Large Programmes, and in continuation of a project started three years
ago, observing time has been awarded at smaller telescopes
with multi-colour photometric instruments (Str\"omgren  photometry: 90cm@Sierra Nevada Observatory (SNO),
1.5m@San Pedro M\'{a}rtir Observatory (SPMO); Geneva photometry: 1.2m
Mercator@Observatorio Roque de los Muchachos (ORM); Johnson
photometry: Konkoly Observatory (KO)). In particular, 18 and 14 consecutive nights have been awarded with the $uvby\beta$ photometers at SPMO and SNO, respectively, in Nov-Dec 2006 and 2007.

\begin{center}
\begin{table}[h]
\caption{\label{logbook} Logbook of the spectroscopic observations, obtained in Jan-Feb 2007 and May-Jul 2007 with the FEROS, SOPHIE and FOCES instruments, dedicated to a selection of  targets of the CoRoT IR01 (Feb-Apr 2007), LRc1 (summer 2007) and LRa1 (winter 2007-2008) runs. The different columns give the target name, its V magnitude, spectral type, variable type, name of the CoRoT run, amount of spectra obtained with FEROS, SOPHIE and FOCES, respectively. The targets indicated in boldface are discussed in this paper. \\}
\centering
\begin{tabular}{llllllll}
\br
target & V & Sp.T. & type & CoRoT run & FEROS & SOPHIE & FOCES  \\ 
\mr
{\bf HD~50844} & 9.09 & A2 & $\delta$ Sct & IR01 & 216 & & \\
HD~50747 & 5.45 & A4 & SB2 & IR01 & 17 & 14 & 6\\
HD~51106 & 7.35 & A3m & SB2 & IR01 & 15 & 14 & 4\\
HD~50846 & 8.43 & B5 & EB & IR01 & 16 & 12 & 4\\ \hline
HD~49434 & 5.74 & F1V & $\gamma$ Dor & LRa1 & 71 & 444 & 75 \\
HD~50209 & 8.36 & B9Ve & Be star & LRa1 & 68 & & \\ \hline
HD~181555 & 7.52 & A5 & $\delta$ Sct & LRc1 & 343 & 66 & 285 \\
HD~174966 & 7.72 & A3 & $\delta$ Sct & LRc1 & & & 119 \\
{\bf HD~180642} & 8.27 & B1.5III & $\beta$ Cep & LRc1 & 213 & 35 & \\
HD~181231 & 9.69 & B9.0V & Be star & LRc1 & 72 & & \\
\br
\end{tabular}
\small{SB2=double-lined spectroscopic binary; EB=eclipsing binary}
\end{table}
\end{center}

In the next sections we report on the preliminary results of the
analysis of the ground-based datasets of the $\delta$ Sct star HD~50844 and the
$\beta$ Cep star HD~180642, and give an outlook towards an interesting candidate CoRoT target, the  hybrid $\delta$ Sct/$\gamma$ Dor star HD~44195.

\section{The $\delta$ Sct star HD~50844}
The $\delta$ Sct star HD~50844 was observed as secondary target around
the solar-like oscillator HD~49933 during CoRoT's IR01 (Feb-Apr
2007). Being a faint target ($V=9.09$), obtaining ground-based
high-resolution spectra with sufficiently high signal-to-noise ratio
(S/N$>$100) proved to be a challenging task. In total 216 \'echelle
spectra were obtained with FEROS/2.2m at ESO, La Silla, during the
nights of 1--10 and 24--28 January 2007 (see \Tref{logbook}).

\begin{figure}
\begin{minipage}{18pc}
\includegraphics[width=7.6cm]{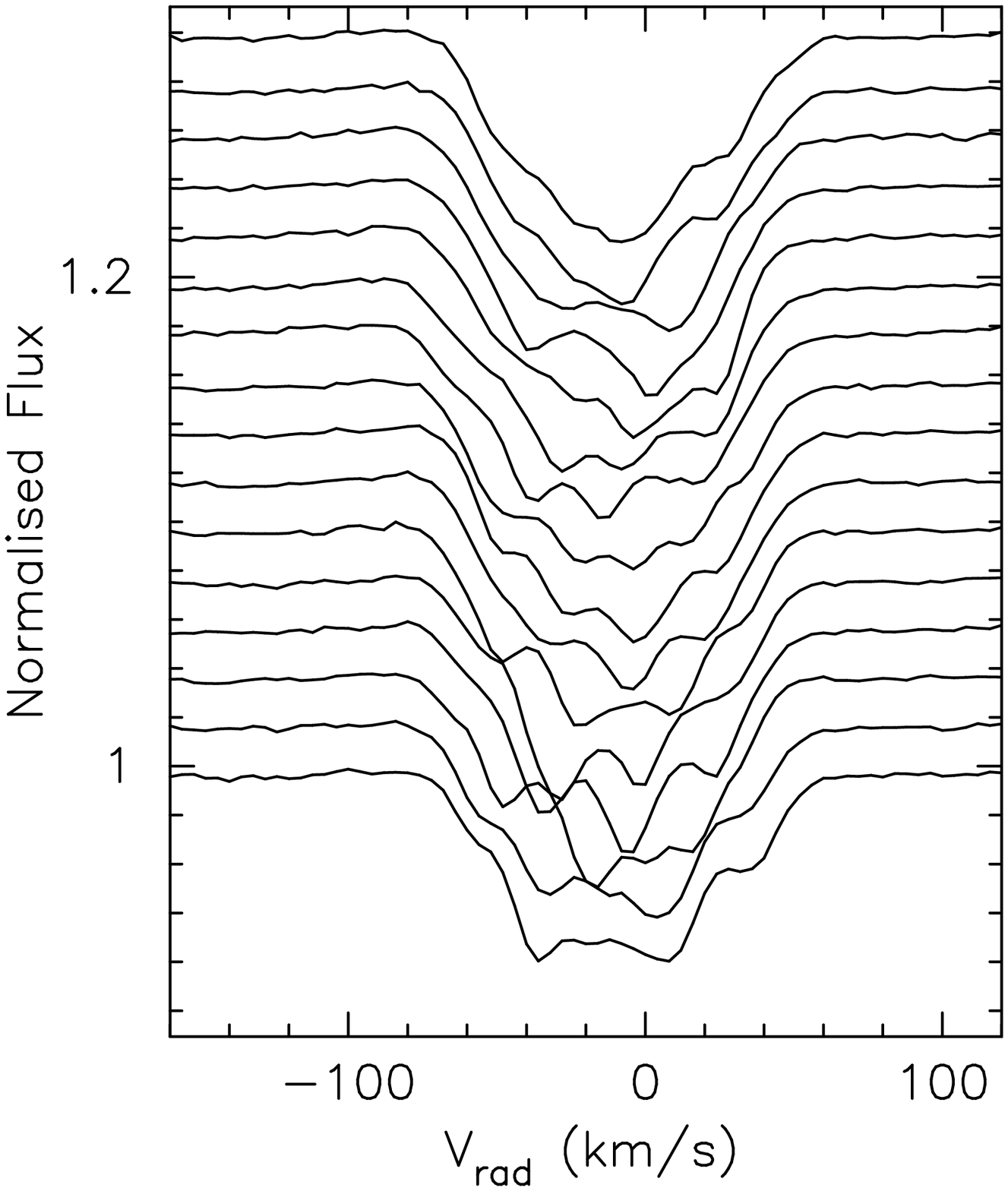}
\caption{\label{spectraHD50844} A set of combined profiles of HD~50844, constructed using the LSD method. The individual profiles are slightly shifted in flux level for visibility reasons.}
\end{minipage}\hspace{2pc}%
\begin{minipage}{17pc}
\includegraphics[width=6.5cm]{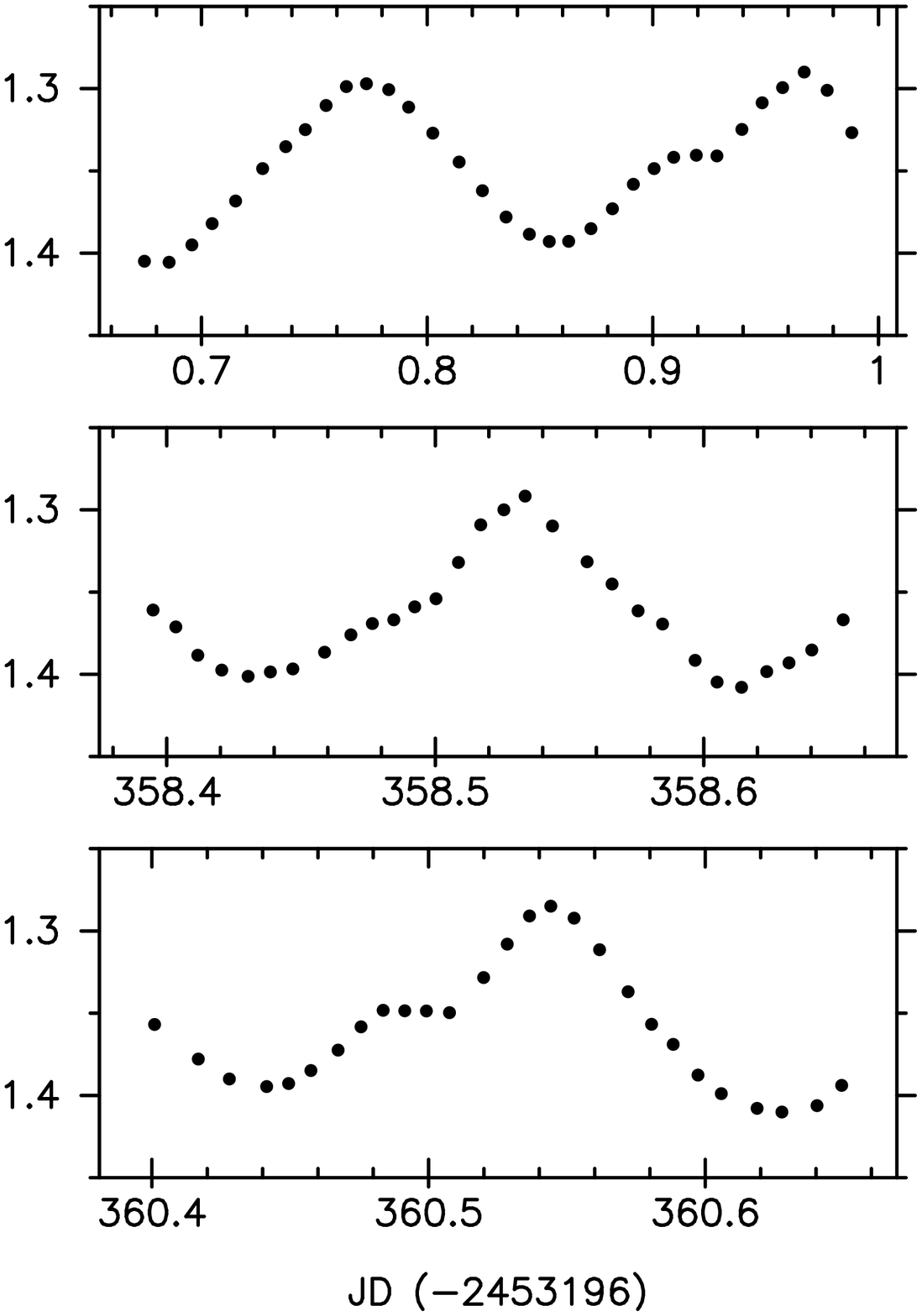}
\caption{\label{standstill} Some lightcurves (Str\"omgren $v$) of HD~180642 obtained at SPMO (top) and SNO (middle and bottom). A non-sinusoidal behaviour, and a 'stillstand', is clearly present. The Julian date is given with respect to JD 2453196.}
\end{minipage}
\end{figure}

Two different teams (INAF-Brera and KULeuven), 
analysed the spectra independently,
using different cross-correlation methods to increase the S/N. One team used the LSD deconvolution method \cite{ref2}, whereby taking
into account more than 1000 lines (Figure \ref{spectraHD50844}), while the other constructed combined profiles from 9 carefully selected lines. 

\begin{figure}
\begin{center}
\hspace{4mm}\includegraphics[width=12cm]{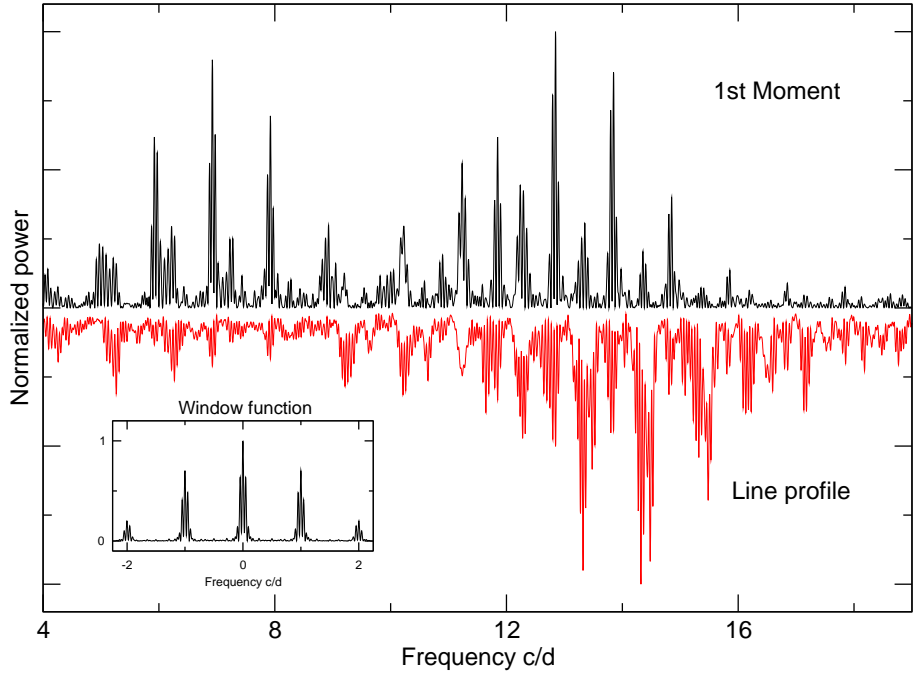}
\end{center}
\caption{\label{powerHD50844} Power spectra of the first moment (top) and pixel-to-pixel intensity variations (bottom) of the combined profiles of HD~50844. The inset shows the associated window function.}
\end{figure}

\begin{figure}
\begin{center}
\hspace{-12mm}\includegraphics[width=12cm]{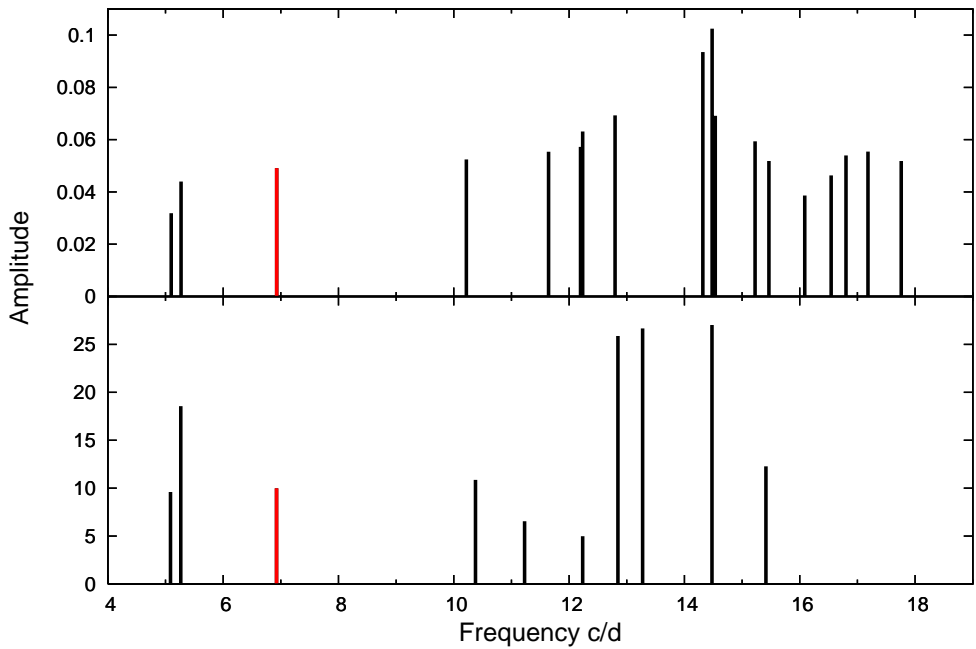}
\end{center}
\caption{\label{frequenciesHD50844} Frequencies detected in the line-profile variations of HD~50844. The upper and lower panel show the results obtained from the pixel-to-pixel analysis and  the moment method (third moment), respectively.
The frequency possibly identified as a radial fundamental mode is indicated in red. The amplitudes are given on a  relative scale.}
\end{figure}

Frequencies were sought for in the line-profile moments \cite{ref3}  and in the
pixel-to-pixel intensity variations \cite{ref4} of the combined profiles
using Fourier analysis and least-squares fitting methods. The
frequency spectrum (Figure \ref{powerHD50844}) shows strong aliasing and
looks particularly difficult to unravel due to closely spaced modes. In the variations of the first moment, more sensitive to the detection of low-degree modes ($\ell<4$), at least nine modes in the range 6--15 d$^{-1}$ are found, with dominating frequencies near 6.9 and 12.8  d$^{-1}$.
 The pixel-to-pixel variations, which allow also the detection of
 high-degree modes, show the presence of at least thirteen frequencies in the range 5--18 d$^{-1}$, with dominating frequency peaks between 14 and
15 d$^{-1}$. 
Only one mode detected in the moment variations was not detected in the pixel-to-pixel variations and therefore can be considered as a 'bona fide' low-degree mode. 
The detected modes are represented in Figure \ref{frequenciesHD50844}.

Preliminary identification of the associated modes, using the phase
 and amplitude variations across the line-profiles \cite{ref4},
 results in prograde modes for the frequencies greater than 13\,d$^{-1}$, and retrograde modes for the frequencies smaller than 6\,d$^{-1}$. The frequency 
6.92\,d$^{-1}$ has the largest amplitude in
 $uvby$ photometry; this fact combined with the preliminary
 spectroscopic analysis suggests  a radial
 fundamental mode.  A more indepth study of the variability of HD~50844
 is currently ongoing.

We expect that the CoRoT time-series of HD~50844, which will be
released at the end of 2007, shall overcome the strong aliasing and
will unravel the complex oscillation pattern with closely spaced
frequencies that we see from the ground. Accurately defined
frequencies and a clear mode-identification are needed for future
asteroseismic modelling of this $\delta$ Sct star.

\section{The $\beta$ Cep star HD~180642}
HD~180642 is to date the only $\beta$ Cep pulsator selected as primary
or secondary target for the CoRoT mission. Consequently, expectations
are high for this sole representative of the $\beta$ Cep class of
pulsators in CoRoT's core programme. In the framework of the
ground-based preparatory programme HD~180642 was monitored in 2004 and
2005 with several photometers (at SPMO, SNO and KO), while it was
already the target of a long-term programme at the Mercator telescope
since 2001. Simultaneously with the CoRoT observations (LRc1, summer
2007), about 240 high-resolution spectra (see \Tref{logbook}) were
obtained with the FEROS and SOPHIE instruments. We report here only on
the analysis of the photometric data and a few exploratory FEROS
spectra that were taken in 2005, for which a logbook is given in
\Tref{photdata}.  The analysis was carried out by teams at KULeuven,
INAF-Brera and Li\`ege.

\begin{table}[h]
\caption{\label{photdata} Logbook of the photometric observations of HD~180642 obtained with P7/Mercator (7 Geneva filters),  the Danish  photometers at SPMO and SNO (Str\"omgren $uvby$) and the 50cm at KO (Johnson V), the available HIPPARCOS (HIP) data, and the spectra obtained with FEROS in 2005. The different columns give an identifier of the telescope or the observatory, the total timespan of the data ($\Delta$T, expressed in days) and the number N of datapoints.} 
\begin{center}
\begin{tabular}{lll|lll}
\br
 & $\Delta$T& N &  & $\Delta$T& N \\
\mr
Mercator & 1184 & 171 & KO & 319 & 69 \\    
SNO & 359 & 235 &	HIP & 1091 & 88 \\  
SPMO & 5.3 & 113 & 	FEROS & 2 & 11\\    
\br
\end{tabular}
\end{center}
\end{table}

In the lightcurves of HD~180642 (Figure \ref{standstill}) the remarkable feature of a 'stillstand' is noticed (cfr. BW Vul \cite{ref5}), which is possibly caused by a shockwave.

\begin{figure}
\begin{center}
{\rotatebox{-90}{\includegraphics[width=10cm]{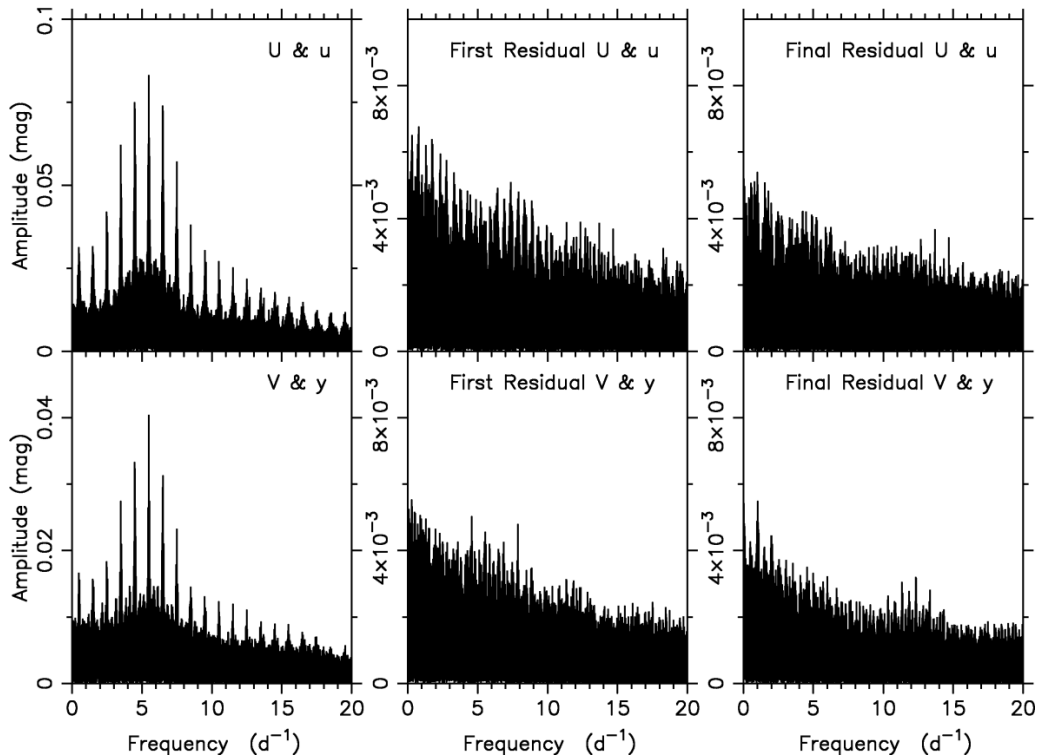}}} 
\end{center}
\caption{\label{frequenciesHD180642} The periodograms for the combined  UV (top left) and violet (bottom left) datasets of HD~180642. The middle panels show the periodograms after prewhitening with $\nu_1$, $2\nu_1$ and $3\nu_1$, while the right panels show the periodograms after subsequent prewhitening with  $\nu_2$, $\nu_3$ and $\nu_4$.}
\end{figure}

To  improve the frequency resolution, we
combined all the UV ($\sim\lambda 3600$ \AA), blue ($\sim\lambda 4500$ \AA) and  violet ($\sim\lambda 5500$ \AA)  data and searched for frequencies using
standard Fourier analysis methods.  We confirm \cite{ref6} the presence of
a dominant frequency $\nu_1 = 5.48695$d$^{-1}$, together with its
harmonics $2\nu_1$ and $3\nu_1$ (Figure \ref{frequenciesHD180642}), which can unambiguously be identified
as a radial mode $\ell=0$, based on the amplitude ratios. By multiplying the normalised periodograms of
the V, $y$ and HIPPARCOS data \cite{ref7}, evidence was found for
$\nu_2=0.3017$ d$^{-1}$ or one of its aliases, which is likely a
g-mode, and possibly two other frequencies between 7 and 8 d$^{-1}$.

From the FEROS spectra listed in \Tref{photdata} we derived abundances
consistent with those of B stars in the solar neighbourhood, but with
a mild nitrogen excess. We obtained $Z=0.0106\pm0.0016$.  Modelling of the radial mode $\nu_1$, using the new solar mixture
\cite{ref8} (which is consistent with the derived abundances), gives, 
contrary to the old 
solar mixture \cite{ref9}, satisfying results when using OP opacities, while OPAL opacities do not predict an unstable mode for a $\sim 10M_{\odot}$ star, such as HD~180642.

The next steps in understanding the variable behaviour of HD~180642
will be the analysis of the recently obtained spectroscopic
time-series (\Tref{logbook}), of new photometric CCD observations from KO, and of the CoRoT timeseries. 
From the complementary
spectroscopic and space data we expect to study the shock wave, to
unravel the frequencies at low amplitude and to identify the g-mode(s),
which will allow to probe the deep interior of HD~180642.

\section{The hybrid $\delta$ Sct/$\gamma$ Dor star HD~44195}
The F0 star HD~44195 ($V=7.56$) is visible in the eyes of CoRoT (LRa not earlier than winter 2009) {\it and} is visible in the MOST viewing zone. For both missions the star still has to be selected.

From line-profile variations HD~44195 is classified as a $\gamma$ Dor
suspect \cite{ref10}, while $\delta$ Sct type variations are detected in
multicolour photometry \cite{ref1}. This makes HD~44195 an interesting
hybrid variable, showing both p- and g-mode oscillations.  As
observational evidence for self-excited hybrid stars is scarse, a
profound study of this star is particularly very interesting.

\begin{figure}
\begin{center}
\includegraphics[width=11cm]{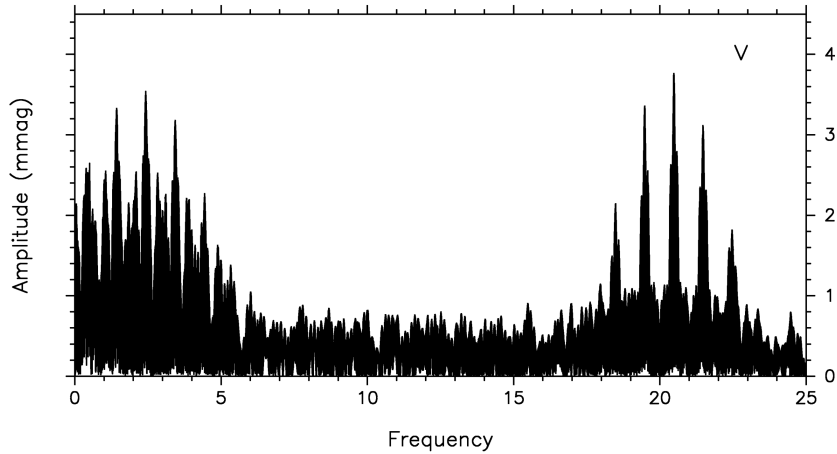}
\end{center}
\caption{\label{HD44195} Frequency spectrum of HD~44195 associated to the $v$ data obtained at SPMO in 2006. The hybrid character becomes clear as frequency peaks near 3 d$^{-1}$  and 21 d$^{-1}$ occur simultaneously, which are typical for $\gamma$ Dor and $\delta$ Sct pulsators, respectively.}
\end{figure}

We observed HD~44195 in Nov-Dec 2006 with the $uvby$ photometers at
SPMO (18 nights) and SNO (14 nights). Similar observations are
scheduled for Nov-Dec 2007. The frequency spectrum associated to the
$v$ magnitude variations shows clearly frequencies in both the
$\gamma$ Dor and $\delta$ Sct region (Figure \ref{HD44195}).

Besides an indepth study of the photometric dataset, we hope to obtain
a spectroscopic timeseries to investigate the genuinety of the hybrid
variability, and to perform a mode-identification with the aim to
constrain the position of HD~44195 in the HR-diagram. Given the
opportunity to probe both the deep layers and the surface of the star,
and to investigate the link between $\gamma$ Dor and $\delta$ Sct
pulsators, we propose the hybrid variable HD~44195 as an excellent target for
both MOST and CoRoT.

\section{Concluding remarks}
We obviously are in a challenging era of asteroseismology. The
preparatory work and the simultaneous ground-based observations of
the CoRoT mission require a huge effort of the asteroseismic
community. Challenging objects are chosen (e.g. faint targets, fast
rotating stars) and in more than one way are we pushing the
limits. The ambitious choice of targets asks for new methodologies and
analysis techniques, and requires a close collaboration between
theoreticians and observers. With joined forces, with shared expertise
and with complementary data from space and from the ground we have
excellent prospects to probe the interiors of stars.

\ack
KU acknowledges financial support from a
\emph{European Community Marie Curie Intra-European Fellowship},
contract number MEIF-CT-2006-024476. This work is supported by the italian ESS project, contract ASI/INAF I/015/07/0, WP 03170, and by the
Research Council of Leuven University under grant GOA/2003/04.

\end{document}